\documentclass[12pt,thmsa]{article}
\usepackage{amsmath}
\usepackage{graphicx}

\begin{document}

\textbf{The Proof that the Standard Transformations of }$\mathbf{E}$\textbf{%
\ and }$\mathbf{B}$\textbf{\ }

\textbf{are not\ the\ Lorentz Transformations. Application to }

\textbf{Motional EMF}\bigskip \bigskip

\qquad Tomislav Ivezi\'{c}

\qquad\textit{Ru%
\mbox
{\it{d}\hspace{-.15em}\rule[1.25ex]{.2em}{.04ex}\hspace{-.05em}}er Bo\v
{s}kovi\'{c} Institute, P.O.B. 180, 10002 Zagreb, Croatia}

\textit{\qquad ivezic@irb.hr\bigskip \medskip }

In this paper it is proved by using the Clifford algebra formalism that the
standard transformations (ST) of the three-dimensional (3D) vectors of the
electric and magnetic fields $\mathbf{E}$ and $\mathbf{B}$ \emph{are not}
the Lorentz transformations (LT) of well-defined quantities from the 4D
spacetime. This difference between the ST and the LT is obtained regardless
of the used algebraic objects (1-vectors or bivectors) for the
representation of the electric and magnetic fields in the usual \emph{%
observer dependent} decompositions of $F$. The LT correctly transform the
\emph{whole} 4D quantity, e.g., $E_{f}=F\cdot \gamma _{0}$, whereas the ST
are the result of the application of the LT only to the part of $E_{f}$,
i.e., to $F$, but leaving $\gamma _{0}$ unchanged. The new decompositions of
$F$ in terms of 4D quantities that are defined without reference frames,
i.e., the absolute quantities, are introduced and discussed. It is shown
that the LT of the 4D quantities representing electric and magnetic fields
correctly describe the motional electromotive force (emf) for all relatively
moving inertial observers, whereas it is not the case with the ST of the 3D $%
\mathbf{E}$ and $\mathbf{B}$. \bigskip


motional emf\bigskip \bigskip

\noindent \textbf{1. INTRODUCTION} \bigskip

\noindent It is generally accepted by physics community that there is an
agreement between the classical electromagnetism and the special relativity
(SR). Such opinion is prevailing in physics already from the Einstein first
paper$^{(1)}$ on SR. The standard transformations (ST) of the 3D vectors of
the electric and magnetic fields, $\mathbf{E}$ and $\mathbf{B}$
respectively, are first derived by Lorentz$^{(2)}$, and independently by
Einstein,$^{(1)}$ and subsequently quoted in almost every textbook and paper
on relativistic electrodynamics. They are considered to be the Lorentz
transformations (LT) of these vectors (see, e.g., Refs. 1,2,3). The same
opinion holds in all usual Clifford algebra formulations of the classical
electromagnetism, e.g., the formulations with Clifford multivectors.$%
^{(4,5,6)}$ In this paper it is proved that the above mentioned ST of $%
\mathbf{E}$ and $\mathbf{B}$ Sec. 3.2 Eq. (\ref{sk1}) (or Sec. 3.3 Eqs. (\ref
{es}), (\ref{bes}) and (\ref{jn1})) \emph{are not} the LT of the 3D $\mathbf{%
E}$ and $\mathbf{B}$. The LT (the active ones) of the electric and magnetic
fields are given by the relations (\ref{nle}) and (\ref{nlb}) Sec. 3.1 (or
Eqs. (\ref{eh}), (\ref{Be}) and (\ref{jn}) Sec. 3.3) and they are quite
different than the ST. The LT always transform the whole 4D quantity
representing the electric or magnetic field, e.g., $E_{f}=F\cdot \gamma _{0}$%
, see Eq. (\ref{nle}) Sec. 3.1, whereas the ST are the result of the
application of the LT only to the part of $E_{f}$, i.e., to $F$, but leaving
$\gamma _{0}$ unchanged, see Eqs. (\ref{ce}) and (\ref{sk1}) Sec. 3.2.
Further in this paper we have presented the new decompositions of $F$ in
terms of well-defined geometric 4D quantities, the 1-vectors of the electric
and magnetic fields $E$ and $B$, as in Eq. (\ref{itf}) Sec. 4, see also Ref.
7, then with the bivectors $E_{Hv}$ and $B_{Hv}$ as in Eq. (\ref{he}) Sec.
4, and with the 1-vector $E_{Jv}$ and the bivector $B_{Jv}$, Eq. (\ref{fje}%
), Sec. 4, which \emph{are all defined without reference frames, }i.e.,\emph{%
\ }they are absolute quantities (AQs).

In the Clifford algebra formalism (as in the tensor formalism) one deals
either with 4D quantities that are defined without reference frames, the
AQs, e.g., the Clifford multivector $F$ (the abstract tensor $F^{ab}$) or,
when some basis has been introduced, with coordinate-based geometric
quantities (CBGQs) that comprise both components and a basis. The SR that
exclusively deals with AQs or, equivalently, with CBGQs, can be called the
invariant SR. The reason for this name is that upon the passive LT any CBGQ
remains unchanged. The invariance of some 4D CBGQ upon the passive LT
reflects the fact that such mathematical, invariant, geometric 4D quantity
represents \emph{the same physical object} for relatively moving observers.
\emph{It is taken in the invariant SR that such 4D geometric quantities are
well-defined not only mathematically but also experimentally, as measurable
quantities with real physical meaning. Thus they have an independent
physical reality. }The invariant SR is discussed in Ref. 8 in the Clifford
algebra formalism and in Refs. 9, 10 in the tensor formalism. It is
explicitly shown in Ref. 10 that the true agreement with experiments that
test SR exists when the theory deals with well-defined 4D quantities, i.e.,
the quantities that are invariant upon the passive LT. The usual ST of the
electric and magnetic fields, the transformations (\ref{ce}), (\ref{B}) and (%
\ref{sk1}), i.e., Eq. (\ref{e7}), Sec. 3.2 (or Eqs. (\ref{es}), (\ref{bes})
and (\ref{jn1}) Sec. 3.3) are typical examples of the ``apparent''
transformations that are first discussed in Refs. 11,12. The ``apparent''
transformations of the spatial distances (the Lorentz contraction) and the
temporal distances (the dilatation of time) are elaborated in detail in
Refs. 9,10, see also Ref. 13. The ``apparent'' transformations relate, in
fact, the quantities from ``3+1'' space \emph{and} time (spatial, or
temporal distances taken separately) and not well-defined 4D quantities.
But, in contrast to the LT of well-defined 4D quantities, \emph{the
``apparent'' transformations do not refer to the same physical object} for
relatively moving observers.

In Secs. 5.1 and 5.2 we have considered the motional emf in two relatively
moving 4D inertial frames of reference using the 3D quantities $\mathbf{E}$
and $\mathbf{B}$ and their ST and the geometric 4D quantities and their LT.
It is shown that the emf obtained by the application of the ST is different
for relatively moving 4D observers. When the geometric 4D quantities and
their LT are used then the emf is always the same; it is independent of the
chosen reference frame and of the chosen system of coordinates in it.

The same proof as here is also presented in the tensor formalism in Ref. 14.
The disagreement between the LT and the ST of the electric and magnetic
fields, that is proved in the Clifford algebra formalism in this paper, and
in the tensor formalism in Ref. 14, is used in Ref. 7 to prove in a
mathematically rigorous manner that, contrary to the general belief, the
usual Maxwell equations with the 3D $\mathbf{E}$ and $\mathbf{B}$ are not
covariant upon the LT but upon the ST. Furthermore, in Ref. 7, the new
Lorentz invariant field equations are presented with well-defined 4D
quantities, the AQs. \bigskip \medskip

\noindent \textbf{2. THE}\ $\gamma _{0}$ - \textbf{SPLIT AND\ THE USUAL
EXPRESSIONS FOR}

$\mathbf{E}$\ \textbf{AND}\ $\mathbf{B}$\ \textbf{IN THE}\ $\gamma _{0}$ -
\textbf{FRAME} \bigskip \medskip

\noindent \textbf{2.1. A Brief} \textbf{Summary} \textbf{of} \textbf{%
Geometric Algebra} \bigskip

First we provide a brief summary of Clifford algebra with multivectors, see,
e.g., Refs. 4,5,6. We write Clifford vectors in lower case ($a$) and general
multivectors (Clifford aggregate) in upper case ($A$). The space of
multivectors is graded and multivectors containing elements of a single
grade, $r$, are termed homogeneous and written $A_{r}.$ The geometric
(Clifford) product is written by simply juxtaposing multivectors $AB$. A
basic operation on multivectors is the degree projection $\left\langle
A\right\rangle _{r}$ which selects from the multivector $A$ its $r-$ vector
part ($0=$ scalar, $1=$ vector, $2=$ bivector, ....). We write the scalar
(grade-$0$) part simply as $\left\langle A\right\rangle .$ The geometric
product of a grade-$r$ multivector $A_{r}$ with a grade-$s$ multivector $%
B_{s}$ decomposes into $A_{r}B_{s}=\left\langle AB\right\rangle _{\
r+s}+\left\langle AB\right\rangle _{\ r+s-2}...+\left\langle AB\right\rangle
_{\ \left| r-s\right| }.$ The inner and outer (or exterior) products are the
lowest-grade and the highest-grade terms respectively of the above series $%
A_{r}\cdot B_{s}\equiv \left\langle AB\right\rangle _{\ \left| r-s\right| },$
and $A_{r}\wedge B_{s}\equiv \left\langle AB\right\rangle _{\ r+s}.$ For
vectors $a$ and $b$ we have $ab=a\cdot b+a\wedge b,$ where $a\cdot b\equiv
(1/2)(ab+ba),$ and $a\wedge b\equiv (1/2)(ab-ba).$ Reversion is an invariant
kind of conjugation, which is defined by $\widetilde{AB}=\widetilde{B}%
\widetilde{A},$ $\widetilde{a}=a,$ for any vector $a$, and it reverses the
order of vectors in any given expression. Any multivector $A$ is a geometric
4D quantity defined without reference frame, i.e., an AQ. When some basis
has been introduced $A$ can be written as a CBGQ comprising both components
and a basis. Usually, Refs. 4,5,6, one introduces the standard basis. The
generators of the spacetime algebra are taken to be four basis vectors $%
\left\{ \gamma _{\mu }\right\} ,\mu =0,...3$ (the standard basis) satisfying
$\gamma _{\mu }\cdot \gamma _{\nu }=\eta _{\mu \nu }=diag(+---).$ This basis
is a right-handed orthonormal frame of vectors in the Minkowski spacetime $%
M^{4}$ with $\gamma _{0}$ in the forward light cone. The $\gamma _{k}$ ($%
k=1,2,3$) are spacelike vectors. The basis vectors $\gamma _{\mu }$ generate
by multiplication a complete basis for the spacetime algebra: $1,\gamma
_{\mu },\gamma _{\mu }\wedge \gamma _{\nu },\gamma _{\mu }\gamma _{5,}\gamma
_{5}$ ($16$ independent elements). $\gamma _{5}$ is the pseudoscalar for the
frame $\left\{ \gamma _{\mu }\right\} .$

We remark that the standard basis corresponds, in fact, to the specific
system of coordinates, i.e., to Einstein's system of coordinates. In the
Einstein system of coordinates the Einstein synchronization$^{(1)}$ of
distant clocks and Cartesian space coordinates $x^{i}$ are used in the
chosen inertial frame of reference. However different systems of coordinates
of an inertial frame of reference are allowed and they are all equivalent in
the description of physical phenomena. For example, in Ref. 9 two very
different, but completely equivalent systems of coordinates, the Einstein
system of coordinates and ``radio'' (``r'') system of coordinates, are
exposed and exploited throughout the paper. The CBGQs representing some 4D
physical quantity in different relatively moving inertial frames of
reference, or in different systems of coordinates in the chosen inertial
frame of reference, are all mathematically equal and thus they are \emph{the
same quantity }for different observers, or in different systems of
coordinates. Then, e.g., the position 1-vector $x$ (a geometric quantity)
can be decomposed in the $S$ and $S^{\prime }$ frames and in the standard
basis $\left\{ \gamma _{\mu }\right\} $ as $x=x^{\mu }\gamma _{\mu
}=x^{\prime \mu }\gamma _{\mu }^{\prime }.$ The primed quantities are the
Lorentz transforms of the unprimed ones. In such interpretation the LT are
considered as passive transformations; both the components and the basis
vectors are transformed but the whole geometric quantity remains unchanged.
Thus we see that \emph{under the passive LT a well-defined quantity on the
4D spacetime, i.e., a CBGQ, is an invariant quantity.}

In the usual Clifford algebra formalism$^{(4,5,6)}$ instead of working only
with such \emph{observer independent quantities} one introduces a space-time
split and the relative vectors. By singling out a particular time-like
direction $\gamma _{0}$ we can get a unique mapping of spacetime into the
even subalgebra of spacetime algebra. For each event $x$ this mapping is
specified by $x\gamma _{0}=ct+\mathbf{x,}$ $\ ct=x\cdot \gamma _{0},\
\mathbf{x}=x\wedge \gamma _{0}$. The set of all position vectors $\mathbf{x}$
is the 3D position space of the observer $\gamma _{0}$ and it is designated
by $P^{3}$. The elements of $P^{3}$ are called \textit{the relative vectors}
(relative to $\gamma _{0})$ and they will be designated in boldface. The
explicit appearance of $\gamma _{0}$ implies that \emph{the space-time split
is observer dependent}. If we consider the position 1-vector $x$ in another
relatively moving inertial frame of reference $S^{\prime }$ (characterized
by $\gamma _{0}^{\prime }$) then the space-time split in $S^{\prime }$ and
in the Einstein system of coordinates is $x\gamma _{0}^{\prime }=ct^{\prime
}+\mathbf{x}^{\prime }$\textbf{.} This $x\gamma _{0}^{\prime }$ is not
obtained by the LT from $x\gamma _{0}.$ (The hypersurface $t^{\prime
}=const. $ is not connected in any way with the hypersurface $t=const.$)
Thence the spatial and the temporal components ($\mathbf{x}$, $t$) of some
geometric 4D quantity ($x$) (and thus the relative vectors as well) are not
physically well-defined quantities. Only their union is physically
well-defined quantity in the 4D spacetime from the invariant SR
viewpoint.\bigskip \medskip

\noindent \textbf{2.2. The\ Usual\ Expressions for }$\mathbf{E}$\textit{\ }%
\textbf{and}\textit{\ }$\mathbf{B}$\textit{\ }\textbf{in the}\textit{\ }$%
\gamma _{0}$\textit{\ }- \textbf{Frame\bigskip }

Let us now see how the space-time split is introduced in the usual Clifford
algebra formalism$^{(4,5)}$ of electromagnetism. The bivector field $F$ is
expressed in terms of the sum of a relative vector $\mathbf{E}$ and a
relative bivector $\gamma _{5}\mathbf{B}$ by making a space-time split in
the $\gamma _{0}$ - frame
\begin{align}
F& =\mathbf{E}_{H}+c\gamma _{5}\mathbf{B}_{H}\mathbf{,\quad E}_{H}=(F\cdot
\gamma _{0})\gamma _{0}=(1/2)(F-\gamma _{0}F\gamma _{0}),  \notag \\
\gamma _{5}\mathbf{B}_{H}& =(1/c)(F\wedge \gamma _{0})\gamma
_{0}=(1/2c)(F+\gamma _{0}F\gamma _{0}).  \label{FB}
\end{align}
(The subscript $H$ is for ``Hestenes.'') Both $\mathbf{E}_{H}$ and $\mathbf{B%
}_{H}$ are, in fact, bivectors. Similarly in Ref. 6 $F$ is decomposed in
terms of 1-vector $\mathbf{E}_{J}$ and a bivector $\mathbf{B}_{J}$ (the
subscript $J$ is for ``Jancewicz'') as
\begin{equation}
F=\gamma _{0}\wedge \mathbf{E}_{J}-c\mathbf{B}_{J},\quad \mathbf{E}%
_{J}=F\cdot \gamma _{0},\ \mathbf{B}_{J}=-(1/c)(F\wedge \gamma _{0})\gamma
_{0}.  \label{J}
\end{equation}
Instead of to use $\mathbf{E}_{H},$ $\mathbf{B}_{H}$ or $\mathbf{E}_{J},$ $%
\mathbf{B}_{J}$ we shall mainly deal (except in Secs. 3.3 and 4) with
simpler but completely equivalent expressions in the $\gamma _{0}$ - frame,
i.e., with 1-vectors that will be denoted as $E_{f}$ and $B_{f}.$ Then
\begin{align}
F& =E_{f}\wedge \gamma _{0}+c(\gamma _{5}B_{f})\cdot \gamma _{0},  \notag \\
E_{f}& =F\cdot \gamma _{0},\ B_{f}=-(1/c)\gamma _{5}(F\wedge \gamma _{0}).
\label{ebg}
\end{align}
All these quantities can be written as CBGQs in the standard basis $\left\{
\gamma _{\mu }\right\} .$ Thus
\begin{equation}
F=(1/2)F^{\mu \nu }\gamma _{\mu }\wedge \gamma _{\nu }=F^{0k}\gamma
_{0}\wedge \gamma _{k}+(1/2)F^{kl}\gamma _{k}\wedge \gamma _{l},\quad
k,l=1,2,3.  \label{EF}
\end{equation}
\begin{align}
E_{f}& =E_{f}^{\mu }\gamma _{\mu }=0\gamma _{0}+F^{k0}\gamma _{k},  \notag \\
B_{f}& =B_{f}^{\mu }\gamma _{\mu }=0\gamma _{0}+(-1/2c)\varepsilon
^{0kli}F_{kl}\gamma _{i}.  \label{gnl}
\end{align}
We see from Eqs. (\ref{EF}) and (\ref{gnl}) that the components of $F$ in
the $\left\{ \gamma _{\mu }\right\} $ basis (i.e., in the Einstein system of
coordinates) give rise to the tensor (components)$\;F^{\mu \nu }=\gamma
^{\nu }\cdot (\gamma ^{\mu }\cdot F)=(\gamma ^{\nu }\wedge \gamma ^{\mu
})\cdot F,$ which, written out as a matrix, has entries
\begin{equation}
E_{f}^{i}=F^{i0},\quad B_{f}^{i}=(-1/2c)\varepsilon ^{0kli}F_{kl}.
\label{sko}
\end{equation}
The relation (\ref{sko}) is nothing else than the standard identification of
the components $F^{\mu \nu }$ with the components of the 3D vectors $\mathbf{%
E}$ and $\mathbf{B,}$ see, e.g., Ref. 3. It is worth noting that \emph{all
expressions with} $\gamma _{0}$, Eq. (\ref{ebg}), \emph{actually refer to
the 3D subspace orthogonal to the specific timelike direction }$\gamma _{0}.$
Really it can be easily checked that $E_{f}\cdot \gamma _{0}=B_{f}\cdot
\gamma _{0}=0,$ which means that they are orthogonal to $\gamma _{0};$ $%
E_{f} $ \emph{and} $B_{f}$ \emph{do not have the temporal components} $%
E_{f}^{0}=B_{f}^{0}=0$. These results, Eq. (\ref{sko}), are quoted in
numerous textbooks and papers treating relativistic electrodynamics, see,
e.g., Ref. 3.

Actually in the usual covariant approaches one forgets about temporal
components $E_{f}^{0}$ and $B_{f}^{0}$ and simply makes the identification
of six independent components of $F^{\mu \nu }$ with three components $E_{i}$
and three components $B_{i}$ according to the relations

\begin{equation}
E_{i}=F^{i0},\quad B_{i}=(-1/2c)\varepsilon _{ikl}F_{kl}.  \label{sko1}
\end{equation}
(The components of the 3D fields $\mathbf{E}$ and $\mathbf{B}$ are written
with lowered (generic) subscripts, since they are not the spatial components
of the 4D quantities. This refers to the third-rank antisymmetric $%
\varepsilon $ tensor too. The super- and subscripts are used only on the
components of the 4D quantities.) Then in the usual covariant approaches,
e.g., Ref. 3, the 3D $\mathbf{E}$ and $\mathbf{B}$, as \emph{geometric
quantities in the 3D space}, are constructed from these six independent
components of $F^{\mu \nu }$ and \emph{the unit 3D vectors }$\mathbf{i},$ $%
\mathbf{j},$ $\mathbf{k,}$ e.g., $\mathbf{E=}F^{10}\mathbf{i}+F^{20}\mathbf{j%
}+F^{30}\mathbf{k}$. (We note that Einstein's fundamental work$^{(15)}$ is
the earliest reference on covariant electrodynamics and on the
identification of some components of $F^{\alpha \beta }$ with the components
of the 3D $\mathbf{E}$ and $\mathbf{B.}$) There is an important difference
between the relations (\ref{sko}) and (\ref{sko1}). As seen from Eqs. (\ref
{ebg})-(\ref{sko1}) $E_{f}$ and $B_{f}$ and their components $E_{f}^{i}$ and
$B_{f}^{i}$ are obtained by a correct mathematical procedure from the
geometric 4D quantities $F$ and $\gamma _{0}$. The components $E_{f}^{i}$
and $B_{f}^{i}$ are multiplied by the unit 1-vectors $\gamma _{i}$ (4D
quantities) to form the geometric 4D quantities $E_{f}$ and $B_{f}$. In such
a treatment the unit 3D vectors\emph{\ }$\mathbf{i},$ $\mathbf{j},$ $\mathbf{%
k,}$ (geometric quantities in the \emph{3D space}) do not appear at any
point. On the other hand the mapping, i.e., the simple identification, Eq. (%
\ref{sko1}), of the components $E_{i}$ and $B_{i}$ with some components of $%
F^{\mu \nu }$ (defined on the 4D spacetime) is not a permissible tensor
operation, i.e., it is not a mathematically correct procedure. The same
holds for the construction of the \emph{3D vectors} $\mathbf{E}$ and $%
\mathbf{B}$ in which the components of the \emph{4D quantity} $F^{\mu \nu }$
are combined with \emph{the unit 3D vectors, }see the second paper in Ref. 7
for the more detailed discussion. Thus it is the relation (\ref{sko}) that
is obtained in a mathematically correct way and not the relation (\ref{sko1}%
).

It has to be noted that the whole procedure that gives Eq. (\ref{sko}) is
made in an inertial frame of reference with the Einstein system of
coordinates. In another system of coordinates that is different than the
Einstein system of coordinates, e.g., differing in the chosen
synchronization (as it is the ``r'' synchronization considered in Ref. 9)
the identification of $E_{f}^{i}$ with $F^{i0},$ as in Eq. (\ref{sko}) (and
also for $B_{f}^{i}$), is impossible and meaningless. In Sec. 4 we shall
present a coordinate-free decomposition of $F$. \bigskip \medskip

\noindent \textbf{3}. \textbf{THE\ PROOF\ THAT THE\ ST\ OF\ }$\mathbf{E}$%
\textbf{\ AND\ }$\mathbf{B}$\ \textbf{ARE\ NOT\ }

\textbf{THE\ LT\bigskip \medskip }

\noindent \textbf{3.1. The Active LT of the Electric and Magnetic Fields}
\textbf{\bigskip }

Let us now explicitly show that the usual transformations of the 3D $\mathbf{%
E}$ and $\mathbf{B}$ are not the LT of quantities that are well-defined on
the 4D spacetime. First we find the correct expressions for the LT (the
active ones) of $E_{f}$ and $B_{f}.$ In the usual Clifford algebra formalism$%
^{(4,5,6)}$ the LT are considered as active transformations; the components
of, e.g., some 1-vector relative to a given inertial frame of reference
(with the standard basis $\left\{ \gamma _{\mu }\right\} $) are transformed
into the components of a new 1-vector relative to the same frame (the basis $%
\left\{ \gamma _{\mu }\right\} $ is not changed). Furthermore the LT are
described with rotors $R,$ $R\widetilde{R}=1,$ in the usual way as $%
p\rightarrow p^{\prime }=Rp\widetilde{R}=p_{\mu }^{\prime }\gamma ^{\mu }.$
But every rotor in spacetime can be written in terms of a bivector as $%
R=e^{\theta /2}.$ For boosts in arbitrary direction
\begin{equation}
R=e^{\theta /2}=(1+\gamma +\gamma \beta \gamma _{0}n)/(2(1+\gamma ))^{1/2},
\label{err}
\end{equation}
$\theta =\alpha \gamma _{0}n,$ $\beta $ is the scalar velocity in units of $%
c $, $\gamma =(1-\beta ^{2})^{-1/2}$, or in terms of an `angle' $\alpha $ we
have $\tanh \alpha =\beta ,$ $\cosh \alpha =\gamma ,$ $\sinh \alpha =\beta
\gamma ,$ and $n$ is not the basis vector but any unit space-like vector
orthogonal to $\gamma _{0};$ $e^{\theta }=\cosh \alpha +\gamma _{0}n\sinh
\alpha .$ One can also express the relationship between the two relatively
moving frames $S$ and $S^{\prime }$ in terms of rotor as $\gamma _{\mu
}^{\prime }=R\gamma _{\mu }\widetilde{R}.$ For boosts in the direction $%
\gamma _{1}$ the rotor $R$ is given by the relation (\ref{err}) with $\gamma
_{1}$ replacing $n$ (all in the standard basis $\left\{ \gamma _{\mu
}\right\} $). Then using Eq. (\ref{gnl}) the transformed $E_{f}^{\prime }$
can be written as
\begin{align}
E_{f}^{\prime }& =R(F\cdot \gamma _{0})\widetilde{R}=R(F^{k0}\gamma _{k})%
\widetilde{R}=E_{f}^{\prime \mu }\gamma _{\mu }=  \notag \\
& =-\beta \gamma E_{f}^{1}\gamma _{0}+\gamma E_{f}^{1}\gamma
_{1}+E_{f}^{2}\gamma _{2}+E_{f}^{3}\gamma _{3},  \label{nle}
\end{align}
what is the usual form for the active LT of the 1-vector $E_{f}=E_{f}^{\mu
}\gamma _{\mu }$. Similarly we find for $B_{f}^{\prime }$
\begin{align}
B_{f}^{\prime }& =R\left[ -(1/c)\gamma _{5}(F\wedge \gamma _{0})\right]
\widetilde{R}=R\left[ (-1/2c)\varepsilon ^{0kli}F_{kl}\gamma _{i}\right]
\widetilde{R}=  \notag \\
& =B_{f}^{\prime \mu }\gamma _{\mu }=-\beta \gamma B_{f}^{1}\gamma
_{0}+\gamma B_{f}^{1}\gamma _{1}+B_{f}^{2}\gamma _{2}+B_{f}^{3}\gamma _{3},
\label{nlb}
\end{align}
what is the familiar form for the active LT of the 1-vector $%
B_{f}=B_{f}^{\mu }\gamma _{\mu }$. It is important to note that $%
E_{f}^{\prime }$ \emph{and} $B_{f}^{\prime }$ \emph{are not orthogonal to} $%
\gamma _{0},$ i.e., \emph{they} \emph{have} \emph{the temporal components} $%
\neq 0.$ They do not belong to the same 3D subspace as $E_{f}$ and $B_{f},$
but they are in the 4D spacetime spanned by the whole standard basis $%
\left\{ \gamma _{\mu }\right\} $. The relations (\ref{nle}) and (\ref{nlb})
imply that the space-time split in the $\gamma _{0}$ - system is not
possible for the transformed $F^{\prime }=RF\widetilde{R}$, i.e., $F^{\prime
}$ cannot be decomposed into $E_{f}^{\prime }$ and $B_{f}^{\prime }$ as $F$
is decomposed in the relation (\ref{ebg}), $F^{\prime }\neq E_{f}^{\prime
}\wedge \gamma _{0}+c(\gamma _{5}B_{f}^{\prime })\cdot \gamma _{0}.$ Notice,
what is very important, that \emph{the components} $E_{f}^{\mu }$ ($%
B_{f}^{\mu }$) from Eq. (\ref{gnl}) \emph{transform upon the active LT again
to the components} $E_{f}^{\prime \mu }$ ($B_{f}^{\prime \mu }$) from Eqs. (%
\ref{nle}) ((\ref{nlb})); \emph{there is no mixing of components}. \emph{Thus%
} \emph{by the active LT} $E_{f}$ \emph{transforms to} $E_{f}^{\prime }$
\emph{and} $B_{f}$ \emph{to }$B_{f}^{\prime }.$ Actually, as we said, this
is the way in which every 1-vector transforms upon the active LT.\bigskip
\medskip

\noindent \textbf{3.2. The ST of the Electric and Magnetic Fields }\bigskip

It is assumed in all standard derivations, e.g., Refs. 15, 3, that one can
again perform the same identification of the transformed components $%
F^{\prime \mu \nu }$ with the components of the 3D $\mathbf{E}^{\prime }$
and $\mathbf{B}^{\prime }$ as in Eq. (\ref{sko1}), i.e.,
\begin{equation}
E_{i}^{\prime }=F^{\prime i0},\quad B_{i}^{\prime }=(-1/2c)\varepsilon
_{ikl}F_{kl}^{\prime },  \label{e7}
\end{equation}
where the same remark about the (generic) subscripts holds also here. As we
said before such simple identification is not a mathematically correct
procedure. However a similar identification is performed in all usual
geometric approaches. Namely instead of making the correct LT of $E_{f}$ and
$B_{f}$ as in Sec. 3.1 it is assumed in all Clifford algebra formalisms,
e.g., Refs. 4,5,6, that the relation (\ref{sko}) with the primed quantities
replacing the unprimed ones, has to be valid for the transformed components
as well. This means that \emph{the ST for} $E_{st}^{\prime }$ \emph{and} $%
B_{st}^{\prime }$ (the subscript $st$ is for standard)\emph{\ are derived
assuming that the quantities obtained by the active LT of} $E_{f}$ \emph{and}
$B_{f}$ \emph{are again in the 3D subspace of the} $\gamma _{0}$ \emph{-}
\emph{observer}. Thus for the transformed $E_{st}^{\prime }$ and $%
B_{st}^{\prime }$ again hold that $E_{st}^{\prime 0}=B_{st}^{\prime 0}=0,$
i.e., that $E_{st}^{\prime }\cdot \gamma _{0}=B_{st}^{\prime }\cdot \gamma
_{0}=0$ as for $E_{f}$ and $B_{f}.$ Thence, in contrast to the correct LT of
$E_{f}$ and $B_{f},$ (\ref{nle}) and (\ref{nlb}) respectively, it is
supposed in the usual derivations$^{(4,5,6)}$ that
\begin{align}
E_{st}^{\prime }& =(RF\widetilde{R})\cdot \gamma _{0}=F^{\prime }\cdot
\gamma _{0}=F^{\prime k0}\gamma _{k}=E_{st}^{\prime k}\gamma _{k}=  \notag \\
& =E_{f}^{1}\gamma _{1}+(\gamma E_{f}^{2}-\beta \gamma cB_{f}^{3})\gamma
_{2}+(\gamma E_{f}^{3}+\beta \gamma cB_{f}^{2})\gamma _{3},  \label{ce}
\end{align}
where $F^{\prime }=RF\widetilde{R}$. Similarly we find for $B_{st}^{\prime }$
\begin{align}
B_{st}^{\prime }& =-(1/c)\gamma _{5}(F^{\prime }\wedge \gamma
_{0})=-(1/2c)\varepsilon ^{0kli}F_{kl}^{\prime }\gamma _{i}=B_{st}^{\prime
i}\gamma _{i}=  \notag \\
& B_{f}^{1}\gamma _{1}+(\gamma B_{f}^{2}+\beta \gamma E_{f}^{3}/c)\gamma
_{2}+(\gamma B_{f}^{3}-\beta \gamma E_{f}^{2}/c)\gamma _{3}.  \label{B}
\end{align}
From the transformations (\ref{ce}) and (\ref{B}) one simply finds the
transformations of the spatial components $E_{st}^{\prime i}$ and $%
B_{st}^{\prime i}$
\begin{equation}
E_{st}^{\prime i}=F^{\prime i0},\quad B_{st}^{\prime i}=(-1/2c)\varepsilon
^{0kli}F_{kl}^{\prime },  \label{sk1}
\end{equation}
which is the relation (\ref{sko}) with the primed quantities. As can be seen
from Eqs. (\ref{ce}), (\ref{B}) and (\ref{sk1}) \emph{the transformations for%
} $E_{st.}^{\prime i}$ \emph{and} $B_{st.}^{\prime i}$\emph{\ correspond to}
\emph{the ST of components of the 3D vectors} $\mathbf{E}$ \emph{and} $%
\mathbf{B}$, Eq. (\ref{e7}), which are quoted in almost every textbook and
paper on relativistic electrodynamics including Refs. 1 and 3. These
relations (\ref{ce}), (\ref{B}) and (\ref{sk1}) are explicitly derived and
given in the Clifford algebra formalism, e.g., in Ref. 4, Space-Time Algebra
(Eq. (18.22)), New Foundations for Classical Mechanics (Ch. 9 Eqs.
(3.51a,b)) and in Ref. 6 (Ch. 7 Eqs. (20a,b)). Notice that, in contrast to
the active LT (\ref{nle}) and (\ref{nlb}), \emph{according to the ST} (\ref
{ce}) \emph{and }(\ref{B}) (\emph{i.e.}, (\ref{sk1})) \emph{the transformed
components} $E_{st}^{\prime i}$ \emph{are expressed by the mixture of
components} $E_{f}^{i}$ \emph{and} $B_{f}^{i},$ \emph{and the same holds for}
$B_{st}^{\prime i}$. In all previous treatments of SR, e.g., Refs. 4,5,6
(and Refs. 1,2,3, Ref. 15) the transformations for $E_{st.}^{\prime i}$ and $%
B_{st.}^{\prime i}$ are considered to be the LT of the 3D electric and
magnetic fields. However our analysis shows that the transformations for $%
E_{st.}^{\prime i}$ and $B_{st.}^{\prime i}$, Eq. (\ref{sk1}), are derived
from the transformations (\ref{ce}) and (\ref{B}), which are not the LT; the
LT are given by the relations (\ref{nle}) and (\ref{nlb}).

The same results can be obtained with the passive LT, either by using a
coordinate-free form of the LT (such one as in Ref. 9), or by using the
standard expressions for the LT in the Einstein system of coordinates from
Ref. 3. The passive LT transform always the whole 4D quantity, basis and
components, leaving the whole quantity unchanged. Thus under the passive LT
the field bivector $F$ as a well-defined 4D quantity remains unchanged,
i.e., $F=(1/2)F^{\mu \nu }\gamma _{\mu }\wedge \gamma _{\nu }=(1/2)F^{\prime
\mu \nu }\gamma _{\mu }^{\prime }\wedge \gamma _{\nu }^{\prime }$ (all
primed quantities are the Lorentz transforms of the unprimed ones). In the
same way it holds that, e.g., $E_{f}^{\mu }\gamma _{\mu }=E_{f}^{\prime \mu
}\gamma _{\mu }^{\prime }$. The invariance of some 4D CBGQ upon the passive
LT is the crucial requirement that must be satisfied by any well-defined 4D
quantity. It reflects the fact that such mathematical, invariant, geometric
4D quantity represents \emph{the same physical object} for relatively moving
observers. The use of CBGQs enables us to have clearly and correctly defined
concept of sameness of a physical system for different observers. Thus in
the invariant SR \emph{such quantity that does not change upon the passive
LT has an independent physical reality, both theoretically and
experimentally. }

However it can be easily shown that $E_{f}^{\mu }\gamma _{\mu }\neq
E_{st}^{\prime \mu }\gamma _{\mu }^{\prime }.$ This means that, e.g., $%
E_{f}^{\mu }\gamma _{\mu }$ and $E_{st.}^{\prime \mu }\gamma _{\mu }^{\prime
}$ \emph{are not the same quantity for observers in} $S$ \emph{and} $%
S^{\prime }.$ As far as relativity is concerned the quantities, e.g., $%
E_{f}^{\mu }\gamma _{\mu }$ and $E_{st.}^{\prime \mu }\gamma _{\mu }^{\prime
},$ are not related to one another. Their identification is the typical case
of \emph{mistaken identity}. The fact that they are measured by two
observers ($\gamma _{0}$ - and $\gamma _{0}^{\prime }$ - observers) does not
mean that relativity has something to do with the problem. The reason is
that observers in the $\gamma _{0}$ - system and in the $\gamma _{0}^{\prime
}$ - system are not looking at the same physical object but at two different
objects. \emph{Every observer makes measurement on its own object and such
measurements are not related by the LT.} Thus from the point of view of the
invariant SR the transformations for $E_{st.}^{\prime i}$ and $%
B_{st.}^{\prime i}$, Eq. (\ref{sk1}), are not the LT of some well-defined 4D
quantities. (This is also exactly proved in the tensor formalism in Ref.
14.) Therefore, contrary to the general belief, it is not true from the
invariant SR viewpoint that, e.g., Ref. 3, Jackson's Classical
Electrodynamics, Sec. 11.10: ''A purely electric or magnetic field in one
coordinate system will appear as a mixture of electric and magnetic fields
in another coordinate frame.''

Let us now examine some derivations of the ST that directly deal with the 3D
$\mathbf{E}$ and $\mathbf{B}$. Such derivation is also presented in
Einstein's fundamental paper$^{(1)}$ and it is discussed in detail in Ref.
9, Sec. 5.3. In Ref. 1 Einstein explicitly worked with the Maxwell equations
written in terms of the 3D $\mathbf{E}$ and $\mathbf{B}$. The main point in
his derivation, that is later taken over in numerous papers and textbooks,
e.g., Ref. 16, Sec. 6.2., is the use of the principle of relativity for the
equations with the 3D quantities. Thus Einstein$^{(1)}$ declares: ''Now the
principle of relativity requires that if the Maxwell-Hertz equations for
empty space hold good in system $K$, they also hold good in system $k$, ..
.'' The requirement that the usual Maxwell equations with the 3D $\mathbf{E}$
and $\mathbf{B}$ have the same form in relatively moving inertial frames led
him to the ST for the components of the 3D $\mathbf{E}$ and $\mathbf{B}$,
Eq. (\ref{e7}). From the invariant SR viewpoint the objection to such type
of the derivation of the ST is that the relatively moving inertial frames
are the 4D frames connected by the LT acting on the 4D spacetime.
Accordingly the principle of relativity necessarily refers to the laws
formulated in the 4D spacetimes. Therefore this principle cannot be taken as
the requirement for the form of the laws written in terms of the 3D
quantities $\mathbf{E}$ and $\mathbf{B}$. As a consequence the derivation of
such type as in Ref. 1 is not in accordance with the symmetry of the 4D
spacetime.

Another type of the derivation of the ST that also uses the 3D quantities is
given, e.g., in Ref. 16, Sec. 6.3. In that derivation$^{(16)}$ it is
supposed that the usual expression for the Lorentz force as a 3D vector (%
\emph{geometric quantity in the 3D space}) must be of the same form in two
relatively moving 4D inertial frames of reference, $\mathbf{F}=q\mathbf{E}+q%
\mathbf{V}\times \mathbf{B}$ and $\mathbf{F}^{\prime }=q\mathbf{E}^{\prime
}+q\mathbf{V}^{\prime }\times \mathbf{B}^{\prime }$, Eqs. (6.42) and (6.43)
in Ref. 16. It is stated in Ref. 16, p. 159: ''It will be assumed that eqns
(6.42) and (6.43) refer to the same act of measurements of the fields, ...''
The ST for the 3D $\mathbf{E}$ and $\mathbf{B}$ are then obtained using the
LT of the 4-force and taking from them only the transformations of the
components of the 3D force $\mathbf{F}$. The objections to such derivation,
from the invariant SR viewpoint, are that (i) the form invariance of the 3D
Lorentz force doesn't follow from any physical law; the principle of
relativity doesn't say anything about the form invariance of the 3D
quantities, (ii) $\mathbf{F}$ is not invariant upon the passive LT which
means, according to the above discussion, that, contrary to the quoted
statement from Ref. 16, $\mathbf{F}$ and $\mathbf{F}^{\prime }$ do not refer
to the same quantity in the 4D spacetime. In addition it will be shown here
in Sec. 5 that the form invariance of the 3D Lorentz force doesn't agree
with the experiments.

Yet another derivation of the ST of the 3D $\mathbf{E}$ and $\mathbf{B}$ is
given in the well-known Purcell's textbook,$^{(17)}$ Sec. 6.7., and it also
discussed in detail in Ref. 9. That derivation uses as an essential part the
Lorentz contraction. But, as already said, it is exactly shown in Ref. 9 and
in the comparison with the standard experiments that test SR, Ref. 10, that
the Lorentz contraction is an ''apparent'' transformation which has nothing
to do with the LT.

\emph{These results (both with the active and the passive LT) entail that
the ST of the 3D vectors} $\mathbf{E}$ \emph{and} $\mathbf{B}$ \emph{are the
''apparent'' transformations and not the LT. Consequently, from the
invariant SR viewpoint, the 3D vectors }$\mathbf{E}$ \emph{and} $\mathbf{B}$
\emph{themselves are not well-defined quantities in the 4D spacetime. }The
same conclusion is achieved in the tensor formalism in Ref. 14.\emph{%
\bigskip \medskip }

\noindent \textbf{3.3. The LT and the ST of}\textit{\textbf{\ }}$\mathbf{E}%
_{H},$ $\mathbf{B}_{H}$ \textbf{and}\textit{\textbf{\ }}$\mathbf{E}_{J},$ $%
\mathbf{B}_{J}$ \bigskip

In this section, for the completness, we shall repeat the proof from Secs.
3.1 and 3.2 but using $\mathbf{E}_{H},$ $\mathbf{B}_{H}$ from Refs. 4, 5 and
$\mathbf{E}_{J},$ $\mathbf{B}_{J}$ from Ref. 6. In Ref. 4,5, as explained in
Sec. 2.2, $F$ is decomposed in terms of bivectors $\mathbf{E}_{H}$ and $%
\mathbf{B}_{H}$, whereas in Ref. 6 $F$ is decomposed in terms of 1-vector $%
\mathbf{E}_{J}$ and a bivector $\mathbf{B}_{J}$. Our aim is to show that the
difference between the ST of the 3D vectors $\mathbf{E}$ and $\mathbf{B}$
and the LT will be obtained regardless of the used algebraic objects for the
representation of the electric and magnetic parts in the decomposition of $F$%
. The correct transformations from the invariant SR viewpoint will be
always, as in Sec. 3.1, simply obtained by applying the LT to the \emph{whole%
} considered 4D algebraic objects. Thus it is unimportant which algebraic
objects represent the electric and magnetic fields. What is important is the
way in which their transformations are derived.

First we present this proof for $\mathbf{E}_{H},$ $\mathbf{B}_{H}$. In Ref.
4,5, as already said in Sec. 2.2, the bivector field $F$ is expressed in
terms of the sum of a relative vector $\mathbf{E}_{H}$ and a relative
bivector $\gamma _{5}\mathbf{B}_{H}$ making a space-time split in the $%
\gamma _{0}$ - frame, Eq. (\ref{FB}); $\mathbf{E}_{H}=(F\cdot \gamma
_{0})\gamma _{0}$ and $\gamma _{5}\mathbf{B}_{H}=(1/c)(F\wedge \gamma
_{0})\gamma _{0}$. All these quantities can be written as CBGQs in the
standard basis $\left\{ \gamma _{\mu }\right\} $. Thus
\begin{equation}
\mathbf{E}_{H}=F^{i0}\gamma _{i}\wedge \gamma _{0},\quad \mathbf{B}%
_{H}=(1/2c)\varepsilon ^{kli0}F_{kl}\gamma _{i}\wedge \gamma _{0}.
\label{aj}
\end{equation}
It is seen from Eq. (\ref{aj}) that both bivectors $\mathbf{E}_{H}$ and $%
\mathbf{B}_{H}$ are parallel to $\gamma _{0}$, that is, it holds that $%
\mathbf{E}_{H}\wedge \gamma _{0}=\mathbf{B}_{H}\wedge \gamma _{0}=0$.
Further we see from Eq. (\ref{aj}) that the components of $\mathbf{E}_{H},$ $%
\mathbf{B}_{H}$ in the $\left\{ \gamma _{\mu }\right\} $ basis (i.e., in the
Einstein system of coordinates) give rise to the tensor (components)$\;(%
\mathbf{E}_{H})^{\mu \nu }=\gamma ^{\nu }\cdot (\gamma ^{\mu }\cdot \mathbf{E%
}_{H})=(\gamma ^{\nu }\wedge \gamma ^{\mu })\cdot \mathbf{E}_{H},$ (and the
same for $(\mathbf{B}_{H})^{\mu \nu }$) which, written out as a matrix, have
entries
\begin{align}
(\mathbf{E}_{H})^{i0}& =F^{i0}=-(\mathbf{E}_{H})^{0i}=E^{i},\quad (\mathbf{E}%
_{H})^{ij}=0,  \notag \\
(\mathbf{B}_{H})^{i0}& =(1/2c)\varepsilon ^{kli0}F_{kl}=-(\mathbf{B}%
_{H})^{0i}=B^{i},\quad (\mathbf{B}_{H})^{ij}=0.  \label{ad}
\end{align}

Using the results from Sec. 3.1 we now apply the active LT to $\mathbf{E}%
_{H} $ and $\mathbf{B}_{H}$ from Eq. (\ref{aj}). For simplicity, as in Secs.
3.1 and 3.2, we again consider boosts in the direction $\gamma _{1}$ for
which the rotor $R$ is given by the relation (\ref{err}) with $\gamma _{1}$
replacing $n.$ Then using Eq. (\ref{aj}) the Lorentz transformed $\mathbf{E}%
_{H}^{\prime }$ can be written as
\begin{align}
\mathbf{E}_{H}^{\prime }& =R[(F\cdot \gamma _{0})\gamma _{0}]\widetilde{R}%
=E^{1}\gamma _{1}\wedge \gamma _{0}+\gamma (E^{2}\gamma _{2}\wedge \gamma
_{0}+  \notag \\
& E^{3}\gamma _{3}\wedge \gamma _{0})-\beta \gamma (E^{2}\gamma _{2}\wedge
\gamma _{1}+E^{3}\gamma _{3}\wedge \gamma _{1}).  \label{eh}
\end{align}
The components $(\mathbf{E}_{H}^{\prime })^{\mu \nu }$ that are different
from zero are $(\mathbf{E}_{H}^{\prime })^{10}=E^{1}$, $(\mathbf{E}%
_{H}^{\prime })^{20}=\gamma E^{2},$ $(\mathbf{E}_{H}^{\prime })^{30}=\gamma
E^{3}$, $(\mathbf{E}_{H}^{\prime })^{12}=\beta \gamma E^{2}$, $(\mathbf{E}%
_{H}^{\prime })^{13}=\beta \gamma E^{3}$. $(\mathbf{E}_{H}^{\prime })^{\mu
\nu }$ is antisymmetric, i.e., $(\mathbf{E}_{H}^{\prime })^{\nu \mu }=-(%
\mathbf{E}_{H}^{\prime })^{\mu \nu }$ and we denoted, as in Eq. (\ref{ad}), $%
E^{i}=F^{i0}$. Similarly we find for $\mathbf{B}_{H}^{\prime }$
\begin{align}
\mathbf{B}_{H}^{\prime }& =R[(-1/c)\gamma _{5}((F\wedge \gamma _{0})\gamma
_{0})]\widetilde{R}=B^{1}\gamma _{1}\wedge \gamma _{0}+  \notag \\
& \gamma (B^{2}\gamma _{2}\wedge \gamma _{0}+B^{3}\gamma _{3}\wedge \gamma
_{0})-\beta \gamma (B^{2}\gamma _{2}\wedge \gamma _{1}+B^{3}\gamma
_{3}\wedge \gamma _{1}).  \label{Be}
\end{align}
The components $(\mathbf{B}_{H}^{\prime })^{\mu \nu }$ that are different
from zero are $(\mathbf{B}_{H}^{\prime })^{10}=B^{1}$, $(\mathbf{B}%
_{H}^{\prime })^{20}=\gamma B^{2},$ $(\mathbf{B}_{H}^{\prime })^{30}=\gamma
B^{3},$ $(\mathbf{B}_{H}^{\prime })^{12}=\beta \gamma B^{2}$, $(\mathbf{B}%
_{H}^{\prime })^{13}=\beta \gamma B^{3}$. $(\mathbf{B}_{H}^{\prime })^{\mu
\nu }$ is antisymmetric, i.e., $(\mathbf{B}_{H}^{\prime })^{\nu \mu }=-(%
\mathbf{B}_{H}^{\prime })^{\mu \nu }$ and we denoted, as in Eq. (\ref{ad}), $%
B^{i}=(1/2c)\varepsilon ^{kli0}F_{kl}$. Both equations (\ref{eh}) and (\ref
{Be}) are the familiar forms for the active LT of bivectors, here $\mathbf{E}%
_{H}$ and $\mathbf{B}_{H}$. It is important to note that $\mathbf{E}%
_{H}^{\prime }$ and $\mathbf{B}_{H}^{\prime }$, in contrast to $\mathbf{E}%
_{H}$ and $\mathbf{B}_{H}$, are not parallel to $\gamma _{0}$, i.e., it
\emph{does not hold} that $\mathbf{E}_{H}^{\prime }\wedge \gamma _{0}=%
\mathbf{B}_{H}^{\prime }\wedge \gamma _{0}=0$ and thus there are $(\mathbf{E}%
_{H}^{\prime })^{ij}\neq 0$ and $(\mathbf{B}_{H}^{\prime })^{ij}\neq 0.$
Further, as in Sec. 3.1, \emph{the components} $(\mathbf{E}_{H})^{\mu \nu }$
($(\mathbf{B}_{H})^{\mu \nu }$) \emph{transform upon the active LT again to
the components} $(\mathbf{E}_{H}^{\prime })^{\mu \nu }$ ($(\mathbf{B}%
_{H}^{\prime })^{\mu \nu }$); \emph{there is no mixing of components}. \emph{%
Thus} \emph{by the active LT} $\mathbf{E}_{H}$ \emph{transforms to} $\mathbf{%
E}_{H}^{\prime }$ \emph{and} $\mathbf{B}_{H}$ \emph{to }$\mathbf{B}%
_{H}^{\prime }.$ Actually, as we said, this is the way in which every
bivector transforms upon the active LT.

In contrast to the correct LT of $\mathbf{E}_{H}$ \emph{and} $\mathbf{B}%
_{H}, $ Eqs. (\ref{eh}) and (\ref{Be}) respectively, it is assumed in the
usual Clifford algebra formalism (Ref. 4, Space-Time Algebra (Eq. (18.22)),
New Foundations for Classical Mechanics (Ch. 9 Eqs. (3.51a,b) and Ref. 5,
(Ch. 7.1.2 Eq. (7.33))) that the relation (\ref{ad}), but with the primed
quantities replacing the unprimed ones, has to be valid for the transformed
components as well. This means that \emph{the ST for }$\mathbf{E}%
_{H,st}^{\prime }$ \emph{and} $\mathbf{B}_{H,st}^{\prime }$ \emph{are
derived assuming that the quantities obtained by the active LT of} $\mathbf{E%
}_{H}$ \emph{and} $\mathbf{B}_{H}$ \emph{are again parallel to} $\gamma _{0}$%
\emph{, i.e., that again holds} $\mathbf{E}_{H}^{\prime }\wedge \gamma _{0}=%
\mathbf{B}_{H}^{\prime }\wedge \gamma _{0}=0$ and consequently that $(%
\mathbf{E}_{H,st}^{\prime })^{ij}=(\mathbf{B}_{H,st}^{\prime })^{ij}=0.$
Thence

\begin{align}
\mathbf{E}_{H,st}^{\prime }& =[(RF\widetilde{R})\cdot \gamma _{0}]\gamma
_{0}=(F^{\prime }\cdot \gamma _{0})\gamma _{0}=E^{1}\gamma _{1}\wedge \gamma
_{0}+  \notag \\
& (\gamma E^{2}-\beta \gamma cB^{3})\gamma _{2}\wedge \gamma _{0}+(\gamma
E^{3}+\beta \gamma cB^{2})\gamma _{3}\wedge \gamma _{0},  \label{es}
\end{align}
where $F^{\prime }=RF\widetilde{R}$. Similarly we find for $\mathbf{B}%
_{H,st}^{\prime }$
\begin{align}
\mathbf{B}_{H,st}^{\prime }& =(-1/c)\gamma _{5}[(F^{\prime }\wedge \gamma
_{0})\gamma _{0})]=B^{1}\gamma _{1}\wedge \gamma _{0}+  \notag \\
& (\gamma B^{2}+\beta \gamma E^{3}/c)\gamma _{2}\wedge \gamma _{0}+(\gamma
B^{3}-\beta \gamma E^{2}/c)\gamma _{3}\wedge \gamma _{0}.  \label{bes}
\end{align}
The relations (\ref{es}) and (\ref{bes}) give the familiar expressions for
the ST of the 3D vectors $\mathbf{E}$ and $\mathbf{B.}$ Now, in contrast to
the correct LT of $\mathbf{E}_{H}$ \emph{and} $\mathbf{B}_{H},$ Eqs. (\ref
{eh}) and (\ref{Be}) respectively, \emph{the components} \emph{of the
transformed }$\mathbf{E}_{H,st}^{\prime }$ \emph{are expressed by the
mixture of components} $E^{i}$ \emph{and} $B^{i},$ \emph{and the same holds
for} $\mathbf{B}_{H,st}^{\prime }$.

The same procedure can be easily applied to the transformations of $\mathbf{E%
}_{J},$ $\mathbf{B}_{J}$ from Ref. 6 and it will lead to the same
fundamental difference between the ST of $\mathbf{E}_{J},$ $\mathbf{B}_{J}$
obtained in Ref. 6 and their correct LT. Again the active LT of $\mathbf{E}%
_{J},$ $\mathbf{B}_{J}$ will be given by
\begin{equation}
\mathbf{E}_{J}^{\prime }=R(F\cdot \gamma _{0})\widetilde{R},\quad \mathbf{B}%
_{J}^{\prime }=R[-(1/c)(F\wedge \gamma _{0})\gamma _{0}]\widetilde{R},
\label{jn}
\end{equation}
whereas the ST from Ref. 6 will follow from
\begin{equation}
\mathbf{E}_{J,st}^{\prime }=(RF\widetilde{R})\cdot \gamma _{0},\quad \mathbf{%
B}_{J,st}^{\prime }=-(1/c)[((RF\widetilde{R})\wedge \gamma _{0})\gamma _{0}].
\label{jn1}
\end{equation}
For brevity the whole discussion will not be done here. Of course the
discussion from Sec. 3.2 regarding the passive LT applies in the same
measure to the results of this section.

It is generally argued in all standard treatments, Refs. 3-6, Ref. 15, of
electromagnetism that the components of the 3D vectors $\mathbf{E}$ and $%
\mathbf{B}$ are the components of $F$ (according to the standard
identification Eq. (\ref{sko}), i.e., Eq. (\ref{sko1}), or Eq. (\ref{ad}))
and that they must transform as such (thus as in Eq. (\ref{sk1}), i.e., Eq. (%
\ref{e7}), or in Eqs. (\ref{es}) and (\ref{bes})) upon the LT. The above
results explicitly show that it is not true. In addition we remark that the
components of $F$ in the chosen system of coordinates are actually
determined by the sources and not by the components of the 3D $\mathbf{E}$
and $\mathbf{B}$. In the recent work$^{(18)}$ I have presented the
formulation of the relativistic electrodynamics (independent of the
reference frame and of the chosen system of coordinates in it) that uses
only the bivector field $F.$ This formulation with $F$ field is a
self-contained, complete and consistent formulation that dispenses with
either electric and magnetic fields or the electromagnetic potentials.
Thence \emph{the} $F$ \emph{field is the primary 4D, geometric, quantity for
the whole electromagnetism and not the 3D} $\mathbf{E}$ \emph{and} $\mathbf{B%
}$. Although it is possible to identify the components of the 3D $\mathbf{E}$
and $\mathbf{B}$ with the components of $F$ (according to Eq. (\ref{sko}) or
Eq. (\ref{ad})) in an arbitrary chosen $\gamma _{0}$ - frame with the $%
\left\{ \gamma _{\mu }\right\} $ basis such identification is meaningless
for the Lorentz transformed $F^{\prime }$. Namely $F$ is a geometric
quantity in the 4D spacetime and when it is written as a CBGQ it contains
both components and a basis. The components are coordinate dependent
quantities depending on the chosen basis and, as explained at the end of Sec
2.2, the standard identification is possible only in the Einstein system of
coordinates. Further it is important to note that the LT always act to the
whole geometric quantity, and thus not to some parts of it (e.g., some
components of $F^{\mu \nu }$). These facts taken together show in another
way too that, contrary to the general belief, it is physically meaningless
to make simple identification of the components of the 3D $\mathbf{E}%
^{\prime }$ and $\mathbf{B}^{\prime }$ with the components of the Lorentz
transformed $F^{\prime }$ (as in Eq. (\ref{sk1}), or in Eqs. (\ref{es}) and (%
\ref{bes}), see also the second paper in Ref. 7).\bigskip \medskip

\noindent \textbf{4. THE\ LORENTZ INVARIANT REPRESENTATIONS }

\textbf{OF\ THE\ ELECTRIC\ AND\ MAGNETIC\ FIELDS}\bigskip

In order to have the electric and magnetic fields defined without reference
frames, i.e., \emph{independent of the chosen reference frame and of the
chosen system of coordinates in it}, thus as AQs, one has to replace $\gamma
_{0}$ (the velocity in units of c of an observer at rest in the $\gamma _{0}$%
-system) in the relation (\ref{ebg}) (and Eqs. (\ref{FB}), (\ref{J}) as
well) with $v$. The velocity $v$, that replaces $\gamma _{0}$, and all other
quantities entering into the relations (\ref{ebg}) (and Eqs. (\ref{FB}), (%
\ref{J}) as well) are all AQs. That velocity $v$ characterizes some general
observer. We can say, as in tensor formalism,$^{(19)}$ that $v$ is the
velocity (1-vector) of a family of observers who measures $E$ and $B$
fields. With such replacement the relation (\ref{ebg}) becomes
\begin{align}
F& =(1/c)E\wedge v+(IB)\cdot v,  \notag \\
E& =(1/c)F\cdot v,\quad B=-(1/c^{2})I(F\wedge v),  \label{itf}
\end{align}
where $I$ is the unit pseudoscalar. ($I$ is defined algebraically without
introducing any reference frame, as in Ref. 20 Sec. 1.2.) It holds that $%
E\cdot v=B\cdot v=0$. Of course \emph{the relations for} $E$ \emph{and }$B$,
Eq. (\ref{itf}),\emph{\ are coordinate-free relations and thus they hold for
any observer.} When some reference frame is chosen with the Einstein system
of coordinates in it and when $v$ is specified to be in the time direction
in that frame, i.e., $v=c\gamma _{0}$, then all results of the classical
electromagnetism are recovered in that frame. Namely we can always select a
particular, but otherwise arbitrary, inertial frame of reference $S,$ the
frame of our ``fiducial'' observers in which $v=c\gamma _{0}$ and
consequently the temporal components of $E_{f}^{\mu }$ and $B_{f}^{\mu }$
are zero (the subscript $f$ is for ``fiducial'' and for this name see Ref.
21). Then in that frame the usual Maxwell equations for the spatial
components $E_{f}^{i}$ and $B_{f}^{i}$ (of $E_{f}^{\mu }$ and $B_{f}^{\mu }$%
) will be fulfilled, see Ref. 7. As a consequence the usual Maxwell
equations can explain all experiments that are performed in one reference
frame. Thus the correspondence principle is simply and naturally satisfied.
However as shown above the temporal components of $E_{f}^{\prime \mu }$ and $%
B_{f}^{\prime \mu }$ are not zero; Eqs. (\ref{nle}) and (\ref{nlb}) are the
correct LT, but it is not the case with Eqs. (\ref{ce}) and (\ref{B}). This
means that the usual Maxwell equations cannot be used for the explanation of
any experiment that tests SR, i.e., in which relatively moving observers
have to compare their data \emph{obtained by measurements on the same
physical object.} However, in contrast to the description of the
electromagnetism with the 3D $\mathbf{E}$ and $\mathbf{B,}$ the description
with $E$ and $B$ is correct not only in that frame but in all other
relatively moving frames and it holds for any permissible choice of
coordinates. It is worth noting that the relations (\ref{itf}) are not the
definitions of $E$ and $B$ but they are the relations that connect two
equivalent formulations of electrodynamics, the formulation with the $F$
field$^{(18)}$ and a new one with the $E$ and $B$ fields. Every of these
formulations is an independent, complete and consistent formulation. For
more detail see Ref. 8 where four equivalent formulations are presented, the
$F$ and $E,$ $B$ - formulations and two new formulations with real and
complex combinations of $E$ and $B$ fields. All four formulations are given
in terms of quantities that are defined without reference frames, i.e., the
AQs. Note however that in the $E,$ $B$ - formulation of electrodynamics in
Ref. 8 the expression for the stress-energy vector $T(v)$ and all quantities
derived from $T(v)$ are written for the special case when $v,$ the velocity
of observers who measure $E$ and $B$ fields is $v=cn$, where $n$ is the unit
normal to a hypersurface through which the flow of energy-momentum ($T(n)$)
is calculated. The more general case with $v\neq n$ will be reported
elsewhere.

In addition, as we have already said, the replacement of $\gamma _{0}$ with $%
v$ in the relations (\ref{FB}) and (\ref{J}) also yields the electric and
magnetic fields defined without reference frames, i.e., as AQs. For
completness, we briefly discuss these cases as well. As explained above the
\emph{observer independent} $F$ field is decomposed, Refs. 4,5, in Eq. (\ref
{FB}) in terms of the \emph{observer dependent quantities,} that is, as the
sum of a relative vector $\mathbf{E}_{H}$ and a relative bivector $\gamma
_{5}\mathbf{B}_{H},$ making the space-time split in the $\gamma _{0}$ -
frame. But, similarly as in Eq. (\ref{itf}) we present here a new
decomposition of $F$ into the bivectors $E_{Hv}$ and $B_{Hv}$, \emph{which
are independent of the chosen reference frame and of the chosen system of
coordinates in it;} they are AQs. We define
\begin{align}
F& =E_{Hv}+cIB_{Hv}\mathbf{,\quad }E_{Hv}=(1/c^{2})(F\cdot v)\wedge v  \notag
\\
B_{Hv}& =-(1/c^{3})I[(F\wedge v)\cdot v],\quad IB_{Hv}=(1/c^{3})(F\wedge
v)\cdot v  \label{he}
\end{align}
(The subscript $Hv$ is for ``Hestenes'' with $v$ and not, as usual, Refs. 4,
5, with $\gamma _{0}$.) Of course, as in Eq. (\ref{itf}), \emph{the velocity}
$v$ \emph{and all other quantities entering into} Eq. (\ref{he}) \emph{are
defined without reference frames, i.e., they are AQs.} Consequently Eq. (\ref
{he}) \emph{holds for any observer.} Similarly when $\gamma _{0}$ is
replaced with $v$ the observer dependent decomposition of $F$ in the
relation (\ref{J}) transforms to the new decomposition of $F$ in terms of
1-vector $E_{Jv}$ and a bivector $B_{Jv}$ \emph{that are all AQs}
\begin{equation}
F=(1/c)v\wedge E_{Jv}-cB_{Jv},\quad E_{Jv}=(1/c)F\cdot v,\
B_{Jv}=-(1/c^{3})(F\wedge v)\cdot v.  \label{fje}
\end{equation}
However, it is worth noting that it is much simpler and, in fact, closer to
the classical formulation of the electromagnetism with the 3D $\mathbf{E}$
and $\mathbf{B}$ to work with the decomposition of $F$ into 1-vectors $E$
and $B$, as in Eq. (\ref{itf}), instead of decomposing $F$ into bivectors $%
E_{Hv}$ and $B_{Hv}$ (\ref{he}), or into the 1-vector $E_{Jv}$ and the
bivector $B_{Jv}$, Eq.(\ref{fje}).

We have not mentioned some other references that refer to the Clifford
algebra formalism and its application to the electrodynamics as are, e.g.,
Ref. 22. The reason is that they use the Clifford algebra formalism with
spinors but, of course, they also consider that the ST of the 3D $\mathbf{E}$
and $\mathbf{B}$, Eq. (\ref{sk1}), are the LT of the electric and magnetic
fields. Thus they also did not notice the fundamental difference between the
LT of the 4D quantities, e.g., $E_{f}^{\prime }$ and $B_{f}^{\prime }$, Eqs.
(\ref{nle}) and (\ref{nlb}), and the ST of the 3D quantities $\mathbf{E}$
and $\mathbf{B}$, Eq. (\ref{sk1}). \bigskip \medskip

\noindent \textbf{5. COMPARISON OF THE ST AND THE\ LT\ FOR\ THE\ }

\textbf{MOTIONAL\ ELECTROMOTIVE\ FORCE}\bigskip \medskip

In this section we shall compare the ST and the LT of the electric and
magnetic fields considering motional electromotive force. \bigskip \medskip

\noindent \textbf{5.1. Motional Electromotive Force with 3D Quantities}%
\bigskip

Let us start with the determination of the electromotive force (emf) using
the 3D quantities, the 3D Lorentz force $\mathbf{F}=q\mathbf{E}+q\mathbf{V}%
\times \mathbf{B}$, the 3D $\mathbf{E}$ and $\mathbf{B}$ and their ST. The
motional emf is produced in an electrical circuit when a circuit or part of
a circuit moves in a magnetic field. The example which will be examined is
from Ref. 16 Sec. 6.4. (see Fig. 6.1. in Ref. 16). We shall also discuss the
problem considered in Ref. 17 Sec. 7.2 (see Figs. 7.2, 7.3 and 7.4 in Ref.
17). These examples are very characteristic for the traditional use of the
ST in the explanation of the electromagnetic phenoma from the point of view
of two relatively moving inertial frames of reference. Different variants of
these examples appear in many standard textbooks (e.g., Ref. 23 Sec. 9-5)
and papers in the educational journals (e.g., Ref. 24).

In Ref. 16 the following problem is considered. In the laboratory system $S$
there is a curved stationary conducting rail and a conducting bar that is
moving through a steady uniform magnetic field $\mathbf{B=-}B_{z}\mathbf{k}$
with velocity $\mathbf{V}$ parallel to the $x$ axis. There is no external
applied electric field in $S$, $\mathbf{E=}0$. Purcell$^{(17)}$ considers
the same problem but without the rail, only a conducting bar is moving in a
steady uniform magnetic field $\mathbf{B}$. Any charge in the bar moves
together with the conductor through the $\mathbf{B}$ field. Since the
electrons are the mobile charge carriers they experience a sideways
deflecting force of magnetic origin given by the expression $q\mathbf{V}%
\times \mathbf{B}$. The motion of charge relative to the bar ceases when a
steady state is settled down. In that state the displaced charges give rise
to an electric field such that, everywhere in the interior of the bar, the
electric force on any charge is equal and opposite to the magnetic force $q%
\mathbf{V}\times \mathbf{B}$. The displaced charges also cause an electric
field outside the bar. Both Rosser$^{(16)}$ (Fig. 6.1. (a)) and Purcell$%
^{(17)}$ (Figs. 7.3 (a) and (b)) sketch the field lines for that external
electric field and they look something like the field lines of separated
positive and negative charges. In order to examine the emf Rosser$^{(16)}$
consider the situation in which the moving conductor slides on the curved
stationary conducting rail making electrical contact with the rail. (A very
similar problem is considered in Ref. 23 Sec. 9-5 where in $S$ there is a
conducting bar, infinitely long and of rectangular cross section, that is
moving through the $\mathbf{B}$ field and a stationary galvanometer with two
sliding contacts that touch the bar on opposite sides.) In the usual
approaches the emf $\varepsilon $ of a complete circuit is defined by means
of the Lorentz force $\mathbf{F}$ that acts on a charge $q$ which is at rest
relative to the section $\mathbf{dl}$ of the circuit
\begin{equation}
\varepsilon =\oint (\mathbf{F/}q)\cdot \mathbf{dl.}  \label{eps}
\end{equation}
In the considered case the emf $\varepsilon $ is determined by the
contribution of the magnetic part of $\mathbf{F}$, i.e., $q\mathbf{V}\times
\mathbf{B}$, as
\begin{equation}
\varepsilon =\int_{o}^{l}VBdy=VBl,  \label{ep1}
\end{equation}
where $l$ is the length of the bar and the bar moves parallel to the $y$
axis.

Some remarks to such usual determination of $\varepsilon $ are at place
already here. The important remark is that it is implicitly assumed in Eqs. (%
\ref{eps}) and (\ref{ep1}) that the integral is taken over the whole circuit
at the same moment of time in $S$, say $t=0$. Further both $\mathbf{F}$ and $%
\mathbf{dl}$ are the 3D vectors that do not transform properly upon the LT
and the emf $\varepsilon $ defined by Eq. (\ref{eps}) is not a Lorentz
scalar. The less important remark is that the field lines are not physical
and the pictures with them actually do not help in understanding physical
phenomena when they are looked from relatively moving frames.

How this ''experiment'' is described in the $S^{\prime }$ frame. In $%
S^{\prime }$ the conducting bar is at rest and the conducting rail moves
with velocity $-\mathbf{V}$. (In order to avoid some possible ambiguities in
the determination of the emf in $S^{\prime }$ we shall slightly modify
Rosser's picture. Namely instead of to take that the whole conducting rail
is in the region of the $\mathbf{B}$ field we suppose that the curved part
of the rail is far outside from the region of the $\mathbf{B}$ field.) The
displacement of charge in the isolated conducting bar must exist in both $S$
and $S^{\prime }$. The usual explanation, e.g., Refs. 16 and 17, is the
following. If in $S$ $\mathbf{E=}0$ and the components of $\mathbf{B}$ are $%
\mathbf{(}0,0\mathbf{,-}B\mathbf{)}$\ then, \emph{according to the ST }the
observer in the $S^{\prime }$ frame sees
\begin{eqnarray}
E_{x}^{\prime } &=&E_{z}^{\prime }=0,E_{y}^{\prime }=-\beta \gamma
cB_{z}=\gamma VB  \notag \\
B_{x}^{\prime } &=&B_{y}^{\prime }=0,B_{z}^{\prime }=-\gamma B,  \label{i1}
\end{eqnarray}
where $\mathbf{\beta }=(V/c)\mathbf{i}$ and $\gamma =(1-\beta ^{2})^{-1/2}$
(compare with, e.g., Eqs. (\ref{ce}) and (\ref{B})). Thence \emph{in} $%
S^{\prime }$ there is not only the magnetic field but \emph{an electric field%
} as well. Then, as stated in Ref. 16 p. 165: ''The electric field $%
E_{y}^{\prime }$ in $\Sigma ^{\prime }$ (our $S^{\prime }$) gives rise to
the separation of charges in the ''stationary'' conductor.'' Purcell$^{(17)}$
again sketch the field lines (see Figs. 7.4 (a) and (b)) which resulted from
the induced electric field $\mathbf{E}^{\mathbf{\prime }}$, uniform
throughout the space, and the field of the surface charge distribution.
Rosser asked the reader (Problem 6.13) to interpret the origin of the
electric field present in $S^{\prime }$. Let us assume that the external
magnetic field in $S$ is due to a permanent magnet at rest in $S$. Then, as
discussed in Ref. 16 Sec. 6.8., Unipolar induction (see also Fig. 6.7. in
Ref. 16), a moving magnet has an electric polarization $\mathbf{P}$ that
gives an electric field outside the moving magnet. In all standard
approaches the polarization $\mathbf{P}$ and the magnetization $\mathbf{M}$
in two relatively moving frames are connected by the same ST as are the $%
\mathbf{E}$ and $\mathbf{B}$ fields, see, e.g., Ref. 16 Eqs. (6.78) and
(6.81), or Ref. 24 Eqs. (18-70) and (18-71). Both in Ref. 16 and Ref. 24 it
is argued that when a permanent magnetization is viewed from a moving frame
it produces an electric moment $\mathbf{P=V\times M}/c^{2}$ which, Ref. 24
p. 337: ''..... \emph{is a consequence of the relativistic definition of
simultaneity}, ..''

It has to be noted already here that \emph{the relativity of simultaneity is
not an intristic relativistic effect but an effect that depends on the
chosen synchronization and every permissible synchronization is only a
convention, see Refs. 9 and 10. The physics must not depend on conventions
which means that, contrary to the general belief, the above usual
explanation for the ST of} $\mathbf{P}$ \emph{and} $\mathbf{M}$ \emph{cannot
be physically correct.} The discussion of electrodynamics of moving media
and the ST of $\mathbf{P}$ and $\mathbf{M}$ together with the comparison
with experiments, e.g., the Wilson and Wilson experiment, will be reported
elsewhere.

Let us proceed to the calculation of $\varepsilon ^{\prime }$ in\emph{\ }$%
S^{\prime }$. The contribution of $B_{z}^{\prime }$ to the emf $\varepsilon $%
, Eq. (\ref{eps}), is zero and only the contribution of $E_{y}^{\prime }$
remains, which is
\begin{equation}
\varepsilon ^{\prime }=\int_{o}^{l}\gamma VBdy=\gamma VBl.  \label{epc}
\end{equation}
Obviously \emph{the emf }$\varepsilon ^{\prime }$, Eq. (\ref{epc}), \emph{in
}$S^{\prime }$ \emph{is not equal to the emf} $\varepsilon $, Eq. (\ref{ep1}%
), \emph{determined} \emph{in} $S;$ $\varepsilon ^{\prime }$ is not much
different from $\varepsilon $ only if $V\ll c$.

We see that the emf obtained by the application of the ST is different for
relatively moving 4D observers which indicates in another way that the ST
are not the correct relativistic transformations, i.e., the LT. \bigskip
\medskip

\noindent \textbf{5.2. Motional Electromotive Force with Geometric 4D
Quantities}\bigskip

Let us now consider the same example as in preceding section but using the
4D geometric quantities.

In the usual Clifford algebra approach to SR$^{(4,5)}$ one makes the
space-time split and writes the Lorentz force $K$ (1-vector) in the Pauli
algebra of $\gamma _{0}$. Since, as we said, this procedure is observer
dependent we express $K$ in terms of AQs, 1-vectors $E$ and $B$, that are
considered in Sec. 4, as
\begin{equation}
K=(q/c)\left[ (1/c)E\wedge v+(IB)\cdot v\right] \cdot u,  \label{KEB}
\end{equation}
see also Ref. 8. (Of course the whole consideration could be equivalently
made using $E_{Hv}$ and $B_{Hv}$ or $E_{Jv}$ and $B_{Jv}$ from Sec. 4.) The
notation is as in Sec. 4 and $u$ is the velocity 1-vector of a charge $q$
(it is defined to be the tangent to its world line). In the general case
when charge and observer have distinct worldlines the Lorentz force $K$ (\ref
{KEB}) can be written as a sum of the $v-\perp $ part $K_{\perp }$ and the $%
v-\parallel $ part $K_{\parallel },$ $K=K_{\perp }+K_{\parallel },$ where
\begin{equation}
K_{\perp }=(q/c^{2})(v\cdot u)E+(q/c)((IB)\cdot v)\cdot u,  \label{Kaok}
\end{equation}
\begin{equation}
K_{\parallel }=(-q/c^{2})(E\cdot u)v,  \label{kapa}
\end{equation}
respectively. Of course $K$, $K_{\perp }$ and $K_{\parallel }$ are all 4D
quantities defined without reference frames, the AQs, and the decomposition
of $K$ is an observer independent decomposition. It can be easily verified
that $K_{\perp }\cdot v=0$ and $K_{\parallel }\wedge v=0.$ Both parts can be
written in the standard basis $\left\{ \gamma _{\mu }\right\} $ as CBGQs
\begin{equation}
K_{\perp }=(q/c^{2})(v^{\nu }u_{\nu })E^{\mu }\gamma _{\mu }+(q/c)\widetilde{%
\varepsilon }_{\ \nu \rho }^{\mu }u^{\nu }B^{\rho }\gamma _{\mu },
\label{kc}
\end{equation}
where $\widetilde{\varepsilon }_{\mu \nu \rho }\equiv \varepsilon _{\lambda
\mu \nu \rho }v^{\lambda }$ is the totally skew-symmetric Levi-Civita
pseudotensor induced on the hypersurface orthogonal to $v$, and
\begin{equation}
K_{\parallel }=(-q/c^{2})(E^{\nu }u_{\nu })v^{\mu }\gamma _{\mu }.
\label{ki}
\end{equation}
Speaking in terms of the prerelativistic notions one can say that in the
approach with the 1-vectors $E$ and $B$ $K_{\perp }$ plays the role of the
usual Lorentz force lying on the 3D hypersurface orthogonal to $v$, while $%
K_{\parallel }$ is related to the work done by the field on the charge.
However \emph{in our invariant formulation of SR only both components
together, Eqs. }(\ref{Kaok})\emph{\ and }(\ref{kapa})\emph{, have physical
meaning and they define the Lorentz force both in the theory and in
experiments. }

Then we define the emf also as an invariant 4D quantity, the Lorentz scalar,
\begin{equation}
emf=\int_{\Gamma }(K/q)\cdot dl,  \label{emfi}
\end{equation}
where $dl$ (1-vector) is the infinitesimal spacetime length and $\Gamma $ is
the spacetime curve. (Note that in Eq. (\ref{emfi}) we deal with the scalar
term of the directed integral and $dl$ is a vector-valued measure and not as
usual a scalar.) Let the observers are at rest in the $S$ frame, $v^{\mu
}=(c,0,0,0)$ whence $E^{0}=B^{0}=0$; the $S$ frame is the rest frame of the
``fiducial'' observers, the $\gamma _{0}$ - frame with the $\left\{ \gamma
_{\mu }\right\} $ basis. Thus the components of the 1-vectors in the $%
\left\{ \gamma _{\mu }\right\} $ basis are $E_{f}^{\mu }=(0,0,0,0)$, $%
B_{f}^{\mu }=(0,0,0,-B)$. As it is said in Sec. 5.1, in the laboratory
system $S$ there is a curved stationary conducting rail and a conducting bar
that is moving with the velocity 1-vector $u$. The components of $u$ and $dl$
in the $\left\{ \gamma _{\mu }\right\} $ basis are $u^{\mu }=(\gamma
c,\gamma V,0,0)$, $dl^{\mu }=(0,0,dl^{2}=dy,0)$. Thence $K_{\parallel }^{a}=0
$, $K_{\perp }^{0}=K_{\perp }^{1}=K_{\perp }^{3}=0$, $K_{\perp }^{2}=\gamma
qVB$. When all quantities in Eq. (\ref{emfi}) are written as CBGQs in the $S$
frame with the $\left\{ \gamma _{\mu }\right\} $ basis we find
\begin{equation}
emf=\int_{0}^{l}\gamma VBdy=\gamma VBl.  \label{f1}
\end{equation}
Since the expression (\ref{emfi}) is independent of the chosen reference
frame and of the chosen system of coordinates in it we shall get the same
result in the relatively moving $S^{\prime }$ frame as well;
\begin{equation}
emf=\int_{\Gamma (in\ S)}(K^{\mu }/q)dl_{\mu }=\int_{\Gamma (in\ S^{\prime
})}(K^{\prime \mu }/q)dl_{\mu }^{\prime }=\gamma VBl.  \label{ef2}
\end{equation}
This can be checked directly performing the LT of all 1-vectors as CBGQs
from $S$ to $S^{\prime }$ including the transformation of $v^{\mu }\gamma
_{\mu }$. ($u^{\prime \mu }=(c,0,0,0)$, $v^{\prime \mu }=(\gamma c,-\gamma
V,0,0)$, $K_{\parallel }^{\prime a}=0$, $K_{\perp }^{\prime 0}=K_{\perp
}^{\prime 1}=K_{\perp }^{\prime 3}=0$, $K_{\perp }^{\prime 2}=K_{\perp
}^{2}=\gamma qVB$, $dl^{\prime \mu }=(0,0,dy,0)$.) Notice that in\emph{\ }$%
S^{\prime }$ the velocity 1-vector $u$ of the conducting bar is different
from zero since it does have the temporal component, which is $=c$. Further
from the viewpoint of the observers in $S^{\prime }$ the velocity 1-vector $v
$ of the ''fiducial'' observers contains not only the temporal component but
also the spatial component. In $S^{\prime }$ the components $E_{f}^{\prime
\mu }$ and $B_{f}^{\prime \mu }$ of the 1-vectors $E$ and $B$ respectively
are obtained by the LT (compare with the transformation of the components in
Eqs. (\ref{nle}) and (\ref{nlb})) and in this particular case they are the
same as $E_{f}^{\mu }$ and $B_{f}^{\mu }$ in $S$; $E_{f}^{\mu }$ transforms
by the LT to $E_{f}^{\prime \mu }$ and also $B_{f}^{\mu }$ to $B_{f}^{\prime
\mu }$. We see that, in contrast to the usual approach with the ST, the LT
do not produce the mixing of the components of the electric and magnetic
fields as the 4D quantities.

Comparing the result, Eq. (\ref{f1}), that is obtained by the use of the
geometric 4D quantities and their LT with the result, Eq. (\ref{ep1}), which
is obtained by the use of the 3D quantities $\mathbf{E}$ and $\mathbf{B}$
and their ST, we reveal that Eq. (\ref{f1}) becomes Eq. (\ref{ep1}) in the
classical limit, i.e., for $V\ll c$. But there is a fundamental difference
between the two approaches: Eq. (\ref{ef2}) shows that Eq. (\ref{f1}) holds
in all relatively moving 4D inertial frames of reference, whereas the
comparison of Eq. (\ref{epc}) and Eq. (\ref{ep1}) shows that Eq. (\ref{ep1})
holds only in the laboratory frame $S$. From the viewpoint of the geometric
approach the agreement with the usual approach exists only in the frame of
the ``fiducial'' observers and when $V\ll c$.

There are many similar examples in the literature and always will be
obtained that there is a fundamental difference between the ST of the 3D $%
\mathbf{E}$ and $\mathbf{B}$ and the LT of the geometric 4D quantities
representing the electric and magnetic fields and that the geometric 4D
approach correctly describes the electromagnetic phenomena in all relatively
moving 4D inertial frames of reference. One important experiment, the
Faraday disk, which leads to the same conclusions, is considered in detail
in Ref. 7. \medskip \bigskip

\noindent \textbf{6. CONCLUSIONS}\bigskip

The whole consideration explicitly shows that the 3D quantities $\mathbf{E}$
and $\mathbf{B}$, their transformations and the equations with them are not
well-defined in the 4D spacetime. More generally, we can conclude that \emph{%
the 3D quantities do not have an independent physical reality in the 4D
spacetime. }Contrary to the general belief we find that \emph{it is not true
from the invariant SR viewpoint that observers in relative motion see
different fields; the transformations, Eqs.} (\ref{ce}), (\ref{B}) \emph{and}
(\ref{sk1}), i.e., \emph{Eq.} (\ref{e7}) (\emph{or Eqs. }(\ref{es}),\emph{\ }%
(\ref{bes}\emph{) and }(\ref{jn1}))\emph{\ are not the correct LT. According
to the correct LT; Eqs.} (\ref{nle}) \emph{and} (\ref{nlb}) (\emph{or Eqs. }(%
\ref{eh}),\emph{\ }(\ref{Be})\emph{\ and }(\ref{jn})) \emph{the electric
field transforms only} \emph{to the electric field and the same holds for
the magnetic field.} The consideration of the motional emf in two relatively
moving 4D inertial frames of reference in Secs. 5.1 and 5.2 completely
justifies the relativistic validity of the geometric approach with the 4D
geometric quantities and with their LT. Thence from the invariant SR
viewpoint the physics must be formulated with 4D AQs as in Eqs. (\ref{itf}),
(\ref{he}) and (\ref{fje}), Eqs. (\ref{KEB}), (\ref{Kaok})\emph{, }(\ref
{kapa}) and (\ref{emfi}), or equivalently with the corresponding 4D CBGQs.
For such formulations of electromagnetism see also Ref. 8, where the
Clifford algebra formalism with multivectors is used, or Refs. 9,10 with the
tensor formalism. The principle of relativity is automatically included in
such theory with well-defined geometric 4D quantities, whereas in the
standard approach to SR$^{(1)}$ the principle of relativity is postulated
outside the mathematical formulation of the theory. The comparison with
experiments from Ref. 10 (and Ref. 7) reveals that the true agreement with
experiments that test SR is achieved when such well-defined geometric 4D
quantities are considered.\bigskip \medskip

\noindent \textbf{ACKNOWLEDGMENTS \bigskip }

\noindent I am grateful to Professor Larry Horwitz for his continuos
interest, support and useful comments. \bigskip \medskip

\noindent \textbf{REFERENCES}\bigskip

\noindent 1. A. Einstein, \textit{Ann. Physik.} \textbf{17}, 891 (1905), tr.
by W. Perrett and G.B.

Jeffery, in \textit{The Principle of Relativity} (Dover, New York).

\noindent 2. H.A. Lorentz, \textit{Proceedings of the Academy of Sciences of
Amsterdam,}

6 (1904), in W. Perrett and G.B. Jeffery, in \textit{The Principle of
Relativity}

(Dover, New York).

\noindent 3. J.D. Jackson, \textit{Classical Electrodynamics} (Wiley, New
York, 1977) 2nd

edn.; L.D. Landau and E.M. Lifshitz, \textit{The Classical Theory of Fields,}

(Pergamon, Oxford, 1979) 4th edn.; C.W. Misner, K.S.Thorne, and J.A.

Wheeler, \textit{Gravitation} (Freeman, San Francisco, 1970).

\noindent 4. D. Hestenes, \textit{Space-Time Algebra }(Gordon and Breach,
New York, 1966);

\textit{Space-Time Calculus; }available at: http://modelingnts.la.
asu.edu/evolution.

html; \textit{New Foundations for Classical Mechanics }(Kluwer Academic

Publishers, Dordrecht, 1999) 2nd. edn.; \textit{Am. J Phys.} \textbf{71},
691 (2003).

\noindent 5. C. Doran, and A. Lasenby, \textit{Geometric algebra for
physicists }(Cambridge

University Press, Cambridge, 2003).

\noindent 6. B. Jancewicz, \textit{Multivectors and Clifford Algebra in
Electrodynamics} (World

Scientific, Singapore, 1989).

\noindent 7. T. Ivezi\'{c}, physics/0311043; physics/0409118.

\noindent 8. T. Ivezi\'{c}, hep-th/0207250; hep-ph/0205277.

\noindent 9. T. Ivezi\'{c}, \textit{Found. Phys.} \textbf{31}, 1139 (2001).

\noindent 10. T. Ivezi\'{c}, \textit{Found. Phys. Lett.} \textbf{15}, 27
(2002); physics/0103026; physics/0101091.

\noindent 11. F. Rohrlich, \textit{Nuovo Cimento\ }\textbf{B} \textbf{45},
76 (1966).

\noindent 12. A. Gamba, \textit{Am. J. Phys.} \textbf{35}, 83 (1967).

\noindent 13. T. Ivezi\'{c}, \textit{Found. Phys. Lett.} \textbf{12}, 105
(1999); \textit{Found. Phys. Lett.} \textbf{12},

507 (1999).

\noindent 14. T. Ivezi\'{c}, \textit{Found. Phys.} \textbf{33} 1339 (2003);
hep-th/0302188.

\noindent 15. A. Einstein, \textit{Ann. Physik} \textbf{49,} 769 (1916), tr.
by W. Perrett and G.B.

Jeffery, in \textit{The Principle of Relativity }(Dover, New York).

\noindent 16. W.G.W. Rosser, \textit{Classical Electromagnetism via
Relativity} (Plenum Press, New

York, 1968)

\noindent 17. E.M. Purcell, \textit{Electricity and Magnetism (}McGraw-Hill,
New York, 1985\textit{)}

2nd\textit{. }edn.

\noindent 18. T. Ivezi\'{c}, physics/0305092.

\noindent 19. R.M. Wald, \textit{General Relativity} (The University of
Chicago Press, Chicago,

1984); M. Ludvigsen, \textit{General Relativity,} \textit{A Geometric
Approach }

(Cambridge University Press, Cambridge, 1999); S. Sonego and

M.A. Abramowicz, J. Math. Phys. \textbf{39}, 3158 (1998); D.A. T. Vanzella,

G.E.A. Matsas, H.W. Crater, Am. J. Phys. \textbf{64}, 1075 (1996).

\noindent 20. D. Hestenes and G. Sobczyk, \textit{Clifford Algebra to
Geometric Calculus}

(Reidel, Dordrecht, 1984).

\noindent 22. H.N. N\'{u}\~{n}ez Y\'{e}pez, A.L. Salas Brito, and C.A.
Vargas, Revista

Mexicana de F\'{i}sica \textbf{34}, 636 (1988).

\noindent 23. W.E. Baylis, \textit{Electrodynamics, a Modern Geometric
Approach} (Birkh\"{a}user,

Boston, 1998); P. Lounesto, \textit{Clifford Algebras and Spinors} (Cambridge

University, Cambridge, 1997).

\noindent 24. W.K.H. Panofsky and M. Phillips, \textit{Classical electricity
and magnetism,}

(Addison-Wesley, Reading, Mass., 1962) 2nd edn.

\end{document}